\documentclass[twocolumn,showpacs,preprintnumbers,amsmath,amssymb]{revtex4}

\usepackage{graphicx}% Include figure files
\usepackage{dcolumn}% Align table columns on decimal point
\usepackage{bm}% bold math
\usepackage{color}
\usepackage{amssymb}
\input epsf
\usepackage{fleqn}

\begin{document}

\title{Kerr-Gauss-Bonnet Black Holes: An Analytical Approximation}

\author{S. Alexeyev}
 \email{alexeyev@sai.msu.ru}
 \affiliation{%
Sternberg Astronomical Institute, Lomonosov Moscow State University\\
Universitetsky Porspekt, 13, Moscow, 119991, Russia
}%
\author{N. Popov}%
\affiliation{%
Computer Center of the Russian Academy of Science\\
Vavilova St., 21, Moscow, 119991, Russia
}%
\author{M. Startseva}
 \affiliation{%
Physics Faculty, Lomonosov Moscow State University\\
Vorobievi Gori, Moscow, 119991, Russia
}%
\author{A. Barrau}%
 \email{Aurelien.Barrau@cern.ch}
 \author{J. Grain}%
 \email{grain@lpsc.in2p3.fr}
\affiliation{%
Laboratoire de Physique Subatomique et de Cosmologie\\
Universit\'e Joseph Fourier, CNRS-IN2P3-INPG\\
53, avenue des Martyrs, 38026 Grenoble cedex, France
}%

\date{December 5, 2007}% 

\begin{abstract}
Gauss-Bonnet gravity provides  one of the most promising frameworks to
study curvature corrections  to  the Einstein action in supersymmetric
string theories, while  avoiding ghosts and keeping second order field
equations.  Although  Schwarzschild-type  solutions  for  Gauss-Bonnet
black holes have been known for long, the Kerr-Gauss-Bonnet metric is
missing. In this paper, a  five  dimensional  Gauss-Bonnet approximation is
analytically derived   for spinning  black  holes and  the
related thermodynamical properties are briefly outlined.
\end{abstract}

\maketitle

\section{Introduction}

In any  attempt to perturbatively  quantize gravity as a field theory,
higher-derivative  interactions  must be included in the action.  Such
terms  also  arise  in  the  effective  low-energy  action  of  string
theories.   Furthermore,  higher-derivative   gravity   theories   are
intrinsically attractive as in many cases  they  display  features  of
renormalizability  and  asymptotic  freedom.  Among  such  approaches,
Lovelock  gravity   \cite{love}   is  especially  interesting  as  the
resulting equations of  motion contain no more than second derivatives
of the  metric, include the self  interaction of gravitation,  and are
free of ghosts when  expanding  around flat space. The four-derivative
Gauss-Bonnet  term  is most probably the dominant  correction  to  the
Einstein-Hilbert action \cite{zie} when considering the  dimensionally
extended  Euler  densities  used  in  the  Lovelock  Lagrangian  which
straightforwardly    generalizes    the     Einstein    approach    in
(4+n)-dimensions. The action therefore reads as:
\begin{eqnarray}
S_{GB} & = & \frac{1}{16\pi G}\int d^{D}x\sqrt{-g}\biggl[-2\Lambda +
R \nonumber \\
& + & \alpha (  R_{\mu\nu\alpha\beta} R^{\mu\nu\alpha\beta}
- 4 R_{\alpha\beta} R^{\alpha\beta} + R^2 )\biggl],
\end{eqnarray}
where $\alpha$ is a  coupling  constant of dimension (length)$^2$, and
$G$ the  $D-$dimensional  Newton's constant defined as $G=1/M_*^{D-2}$
in terms of the  fundamental  Planck scale $M_*$. Gauss-Bonnet gravity
was shown to exhibit a very rich phenomenology in cosmology (see, {\it
e.g.}, \cite{char}  and references therein), high-energy physics (see,
{\it e.g.}, \cite{barr}  and references therein) and black hole theory
(see,  {\it  e.g.},  \cite{alex}  and  references  therein).  It  also
provides  interesting   solutions   to   the   dark   energy   problem
\cite{Nojiri},   offers   a   promising   framework   for    inflation
\cite{lidsey,ferrer},    allows    useful    modification    of    the
Randall-Sundrum  model   \cite{kim}   and,   of  course,  solves  most
divergences  associated  with the endpoint of the Hawking  evaporation
process \cite{alex2}.

Either in D-dimensions or in  4-dimensions  with  a dilatonic coupling
(required to make the Gauss-Bonnet term dynamical), Gauss-Bonnet black
holes and their rich thermodynamical properties \cite{simon} have only
been studied in the non-spinning ({\it i.e.} Schwarzschild-like) case.
Although some  general features can be  derived in this  framework, it
remains  mostly  unrealistic  as  both astrophysical black  holes  and
microscopic    black    holes    possibly    formed    at    colliders
\cite{banks,dimo,gidd}  are   expected  to  be  rotating  ({\it  i.e.}
Kerr-like).  Of  course,  the  latter  --  which  should  be copiously
produced at the {\it Large Hadron  Collider} if the Planck scale is in
the  TeV  range  as  predicted  by  some large extra-dimension  models
\cite{ADD}-- are  especially  interesting  for Gauss-Bonnet gravity as
they  could  be  observed  in  the  high-curvature  region  of General
Relativity  and  allow a direct measurement of  the  related  coupling
constant \cite{barr}. The range of impact  parameters corresponding to
the formation of a non-rotating black hole being of zero  measure, the
Schwarzschild  or  Schwarzschild-Gauss-Bonnet   solutions  are  mostly
irrelevant. This is  also  of experimental  importance  as only a  few
quanta  should be  emitted  by those light  black  holes, evading  the
Gibbons \cite{gib} and  Page  \cite{pag} arguments usually pointed out
to neglect the angular momentum of primordial black holes.

It  should  be  underlined  that  D-dimensional  spinning  black  hole
solutions are  anyway  very  important  within  different  theoretical
frameworks ({\it  e.g.},  in conservation law studies) \cite{petrov1}.
Thanks  to   perturbation   theory,   several   attempts   were   made
\cite{petrov2} to  derive the solution.  In the following, we focus on
an analytical approach.

\section{5D Solution}

To  investigate  the detailed properties of black  holes  in  Lovelock
gravity, it is therefore mandatory to  derive  the  general  solution,
{\it  i.e} the  metric  for the spinning  case.  Unlike the  numerical
attempts that  were  presented  in  \cite{alexpopov}  for  degenerated
angular momenta, the present paper focuses on the exact solution  in 5
dimensions.

Einstein  equations   in  Gauss-Bonnet  gravity  with  a  cosmological
constant $\Lambda$ read as
\begin{eqnarray}
&& R_{\mu\nu} - \frac{1}{2} g_{\mu\nu} R =
\Lambda g_{\mu\nu} \nonumber \\
& + & \alpha \Biggl[ \frac{1}{2} g_{\mu\nu} \biggl(
R_{\mu\nu\alpha\beta} R^{\mu\nu\alpha\beta}
- 4 R_{\alpha\beta} R^{\alpha\beta} + R^2 \biggr) \nonumber \\
& - & 2 R R_{\mu\nu}
+ 4 R_{\mu\gamma} R^{\gamma}_{\nu}
+ 4 R_{\gamma\delta} R^{\gamma\delta}_{\mu\nu} \nonumber \\
& - & 2 R_{\mu\gamma\delta\lambda} R_{\nu}^{\gamma\delta\lambda}
\Biggr],
\end{eqnarray}
and the  5-dimensional metric in the spherically-symmertic Kerr-Schild
type can be written as
\begin{eqnarray}\label{metric}
ds^2 & = & dt^2 - dr^2 -
(r^2+a^2)\sin^2\theta d\phi_1^2 \nonumber \\
& - & (r^2+b^2) \cos^2\theta d\phi_2^2
- \rho^2 d\theta^2 \nonumber \\
& - & 2 dr \Biggl( a \sin^2\theta d\phi_1
+ b \cos^2\theta d\phi_2 \Biggr) \nonumber \\
& - & \beta \Biggl( dt - dr
- a \sin^2\theta d\phi_1 \nonumber \\
& - & b \cos^2\theta d\phi_2 \Biggr)^2,
\end{eqnarray}
where $\rho^2  =  r^2  +  a^2  \cos^2\theta
+  b^2  \sin^2\theta$ and unknown function $\beta = \beta(r, \theta)$.

The $\theta\theta$ component of Einstein equations reads :
\begin{equation}
A \beta'' + B \beta'^2 + C\beta' + D \beta + E = 0,
\end{equation}
where
\begin{eqnarray*}
A & = & r \rho^2 (4 \alpha \beta - \rho^2) \\
B & = & 4 \alpha r \rho^2 \\
C & = & 2 \Biggl[ 4 \alpha \beta (\rho^2 -r^2)
- \rho^2 (\rho^2 + r^2) \Biggr] \\
D & = & 2 r ( 2 r^2 - 3 \rho^2) \\
E & = & 2 r \Lambda \rho^4.
\end{eqnarray*}

This equation can be split into  2  relations  depending  respectively
only upon $\beta$ and $z\equiv  \beta  \beta'$  as independent unknown
functions. It  is then possible to  introduce a new  function $f(r,c)$
where $c  = a^2 \cos^2\theta + b^2 \sin^2\theta$  so that the equations
are equivalent to the following system:
\begin{eqnarray}
&& \beta'' + 2 (\frac{\rho^2 + r^2}{r \rho^2} \beta'
- \frac{2r^2-3\rho^2}{\rho^4} \beta
-  \Lambda) \nonumber \\
& - & \frac{f(r,c)}{r \rho^4} =0 \\
&& z' + 2 \frac{\rho^2 - r^2}{r\rho^2} z
-\frac{1}{2} \frac{f(r,c)}{\alpha r \rho^2} =0.
\end{eqnarray}

Introducing the new function $p(r,c)$ via the transformation
\begin{eqnarray}
f(r,c) = \frac{\rho^4}{r} \frac{\partial p(r,c)}{\partial r},
\end{eqnarray}
the second equation can be solved
(with $p_r \equiv \partial p(r,c) /\partial r$), leading to~:
\begin{eqnarray}
z = \frac{1}{2} \frac{(\int p_r dr + 2 C_{21} \alpha)(r^2 + c^2)}
{\alpha r^2} = (\beta \beta'),
\end{eqnarray}
where $C_{ij}$ are constants of  integration  in  the $i$-th equation.
This equation can be integrated to obtain:
\begin{eqnarray}
\beta^2 = \frac{1}{\alpha} \int \Biggl(
p \frac{r^2+c^2}{r^2}\Biggr) dr
+ 2 C_{21} \frac{r^2 - c^2}{r} + C_{20}.
\end{eqnarray}

The first equation results in
\begin{eqnarray}
\beta & = & \Biggl( C_{12} r - C_{11} (r^2 -c^2) \nonumber \\
& - & r \int \frac{(p_r - 2 r^2 \Lambda)(r^2
- c^2 )}{r} dr  \nonumber \\
& + & (r^2+c^2) \int (p_r
+ 2 \Lambda r^2) dr \Biggr)\frac{1}{r (r^2 + c^2)}
\end{eqnarray}
where a simple integration by parts
\begin{eqnarray}
\int p_r \frac{r^2-c^2}{r} dr = p \frac{r^2-c^2}{r}
- \int p \frac{r^2+c^2}{r^2} dr
\end{eqnarray}
leads to:
\begin{eqnarray}
&& \beta r (r^2+c^2) = c_{12} r + C_{11} (r^2-c^2) \nonumber \\
& + & r \int \Biggl( p \frac{r^2 + c^2}{r^2}\Biggr) dr
+ \frac{\Lambda r^3}{6} (r^2+c^2).
\end{eqnarray}

As the same integral combination
\begin{eqnarray}
Q = \int \Biggl( p \frac{r^2+c^2}{r^2}\Biggr) dr .
\end{eqnarray}
is involved, the system leads to the quadratic equation:
\begin{eqnarray}
&& \alpha \beta^2 - (r^2+c^2) \beta \nonumber \\
& + & \Biggl( C_{32} + C_{31} \frac{r^2 - c^2}{r} \nonumber \\
& + & \frac{\Lambda r^2}{6} (r^2+2c^2) \Biggr) = 0
\end{eqnarray}
where  $C_{3i}$  are   new   integration  constants  obtained  from  a
combination of $C_{2i}$ and $C_{1i}$.

Taking into account  the asymptotes at infinity (and therefore finding
the values of the integration constants, $M$ being the ADM mass), this
leads to
\begin{equation}\label{betalast}
\beta = \frac{\rho^2 \pm \sqrt{\rho^4 - 4 \alpha M
- \frac{2}{3} \alpha \Lambda r^2 (2 \rho^2 - r^2)}}{2 \alpha}
\end{equation}
where the  ``-'' branch should be chosen  so as  to recover the  usual
Kerr solution  in the limit  $\alpha \to  0$. In case  of a  vanishing
rotation ($a=b=0$) the obtained solution corresponds  to one suggested
in Ref. \cite{boulware}.  When used in the metric (\ref{metric}), this
leads to the exact Kerr - Gauss - Bonnet - (anti) - deSitter solutions
of Einstein equations. As only the  $\theta\theta$  component  of  the
field equations was used to derive this result, the compatibility with
the other components was carefully checked. Although the equations are
far  too  intricate to allow for analytical investigations,  numerical
results show that they are indeed fulfilled.

\section{Transformation to the Boyer - Linguist form}

To  obtain the value  of  horizon  radius  $r_h$ it  is  necessary  to
transform  the  metric (\ref{metric}) back  to   the
Boyer-Linguist  form with:
\begin{eqnarray*}
dt' & = & A dt + B dr + C d\theta, \\
d\phi_1' & = & D d \phi_1 + E dr + F d\theta, \\
d\phi_2' & = & D d\phi_2 + H dr + F d\theta.
\end{eqnarray*}
Taking  into  account that the processes relevant for  thermodynamical
investigations  take  place  in  the  surroundings   of  the  horizon,
$M/\rho^2$ can be  considered  as  a small parameter and  used  for  a
Taylor expansion of $\beta$ as
\begin{eqnarray*}
\beta \approx \frac{M}{\rho^2} + \frac{8  M^2
\alpha}{\rho^6}.
\end{eqnarray*}
The  Boyer  -  Linguist  parameterization  imposes,   as  a  necessary
condition, vanishing coefficients for  non-diagonal  components except
for  $dt d\phi_1$  and  $dt  d\phi_2$.  This leads  to a  system  of 8
equations  with  8   variables.   The  solutions  are  explicitly  the
Boyer-Linguist parameterization of the Kerr-Gauss-Bonnet metric.

Solving those equations (without substituting the direct expression of
$\rho(r, \theta)$), one obtains that:

i) all the coefficients before the components $d\theta  dx$, $x$ being
an arbitrary coordinate vanish automatically, as in the classical Kerr
case~;

ii) The coefficients $A$ and $D$  could be set equal to $1$ to recover
the classical case;

iii) Other coefficients are:
\begin{eqnarray*}
B & = & B_1 /B_2 \\
E & = & E_1 /E_2 \\
H & = & H_1 /H_2 \\
\end{eqnarray*}
where
\begin{eqnarray*}
B_1 & = & - \rho^6 a (r^2 + b^2) \\
B_2 & = &
+ r^4 \rho^6
+ 8 M^2 \alpha a^2 \cos^2\theta r^2
- M \rho^4 a^2 \cos^2\theta r^2 \\
&& + \rho^6 b^2 r^2
+ \rho^6 b^2 a^2
- M \rho^4 r^4 \\
&& + r^2 \rho^6 a^2
+ 8 M^2 \alpha r^4
- M \rho^4 b^2 r^2 \\
&& + 8 M^2 \alpha b^2 r^2
+ M \rho^4 b^2 \cos^2\theta r^2 \\
&& - 8 M^2 \alpha b^2 \cos^2\theta r^2, \\
E_1 & = & -(a^2+r^2) \rho^6 b \\
E_2 & = &
r^4 \rho^6
+ 8 M^2 \alpha a^2 \cos^2\theta r^2
- M \rho^4 a^2 \cos^2\theta r^2 \\
&& + \rho^6 b^2 r^2
+ \rho^6 b^2 a^2
- M \rho^4 r^4 \\
&& + r^2 \rho^6 a^2
+ 8 M^2 \alpha r^4
- M \rho^4 b^2 r^2 \\
&& + 8 M^2 \alpha b^2 r^2
+ M \rho^4 b^2 cos^2\theta r^2 \\
&& -8 M^2 \alpha b^2 \cos^2\theta r^2, \\
H_1 & = &
+ M r^2 (8 M \alpha a^2 \cos^2\theta
- \rho^4 a^2 \cos^2\theta
- \rho^4 r^2 \\
&& +8 M \alpha r^2
- \rho^4 b^2
+ 8 M \alpha b^2 \\
&& + \rho^4 b^2 \cos^2\theta
- 8 M \alpha b^2 \cos^2\theta) \\
H_2 & = &
+ r^4 \rho^6
+ 8 M^2 \alpha a^2 \cos^2\theta r^2
- M \rho^4 a^2 \cos^2\theta r^2 \\
&& + \rho^6 b^2 r^2
+ \rho^6 b^2 a^2
- M \rho^4 r^4 \\
&& + r^2 \rho^6 a^2
+ 8 M^2 \alpha r^4
- M \rho^4 b^2 r^2 \\
&& + 8 M^2 \alpha b^2 r^2
+ M \rho^4 b^2 cos^2\theta r^2 \\
&& - 8 M^2 \alpha b^2 \cos^2\theta r^2.
\end{eqnarray*}

After substituting those coefficients in the metric (\ref{metric}) and
some algebra, the metric becomes:
\begin{eqnarray}\label{metric2}
&& ds^2 = \\
& + & dt^2 - (r^2+a^2) \cos^2\theta d\phi_1^2
- (r^2+b^2) \sin^2\theta d\phi_2^2 \nonumber \\
& - & \rho^2 d\theta^2 - \Biggl( \frac{M}{\rho^2}
+ \frac{8 M^2 \alpha}{\rho^6}\Biggr) \Biggl( dt \nonumber \\
& + & a \sin^2\theta d\phi_1
+ b \cos^2\theta d\phi_2 \Biggr)^2 \nonumber \\
& - & \frac{\Phi}{\rho^6 \Biggl(
(r^2 + a^2) (r^2+b^2)
- r^2 \rho^2(\frac{M}{\rho^2}
+ \frac{8M^2 \alpha}{\rho^6})\Biggr)}
dr^2
\nonumber
\end{eqnarray}
where  $\Phi$  is a  coefficient  whose value  is  irrelevant as  this
investigation requires only the denominator of the last term ($g_{11}$
component of the metric), which is:
\begin{eqnarray}\label{g11}
\rho^6 \Biggl( (r^2 + a^2)(r^2 + b^2)
- r^2 \rho^2 \beta \Biggr)
\end{eqnarray}

\section{Thermodynamical properties}

For the investigation of the black hole topology, one has to study the
singular points of the metric component $g_{11}$, {\it i.e.} study the
zeros of the expression (\ref{g11}):
\begin{equation}\label{g11a}
(r^2+a^2) (r^2+b^2) - r^2 \rho^2 \beta = 0 .
\end{equation}
This  is an 8th  order  equation  in  $\rho$ when  $\beta$  is  Taylor
expanded at the lowest order in $\alpha$. As show in \cite{labbe}, the
cosmological constant can change the temperature. In the following, we
restrict our study to the $\Lambda=0$ case. Using the value of $\beta$
given in (\ref{betalast}), one obtains:
\begin{eqnarray}\label{M}
M = \frac{M^*}{4 \alpha r^4 \rho^4}
\end{eqnarray}
where
\begin{eqnarray*}
M^* & = &
r_+^4 \rho^8
- 4 \alpha (r_+^2 + a^2)^2 (r_+^2 + b^2)^2 \\
&& + 4 \alpha r_+^2 \rho^4 (r_+^2 + a^2) (r_+^2 + b^2)
+ r_+^8 \rho^4 
\end{eqnarray*}
and $r_+$ is horizon radius.
When $\alpha \to 0$ this leads to the usual Kerr case.

It should  be underlined that  the angle variable $\theta$ is included
in the  expression (\ref{M}) (as $\rho^2 = r_+^2  + a^2 \cos^2\theta +
b^2 \sin^2\theta$), which indicates  a  good choice of coordinates. To
remove this dependence one has  to  set $\theta = \pi/4$. This  allows
computing  the  temperature, which requires the surface gravity  given
by:
\begin{eqnarray}
\kappa^2 = -\frac{1}{4}g^{tt}g^{ij}(\partial_{i}g_{tt})
(\partial_{j}g_{tt})|_{r=r_+}.
\end{eqnarray}
In the considered case, this leads to
\begin{eqnarray}\label{kapa}
\kappa = -\frac{1}{4}(1 + \beta)[g^{rr}(\partial_r
\beta)^2 + g^{\theta \theta}(\partial_{\theta} \beta)^2] |_{r=r_+}.
\end{eqnarray}
After the substituting all the values, this formula becomes
\begin{eqnarray}
\kappa =  - \frac {1}{4} (1 + \beta) \Biggl(
\frac{\kappa_1}{\kappa_2}
- \frac {(\frac{\partial}{\partial \theta}\beta)^2}
{\rho^2} \Biggr),
\end{eqnarray}
where
\begin{eqnarray*}
\kappa_1 & = &
( \beta r^2 \cos^2\theta (a^2 - b^2)
- ( r^2 + a^2 ) ( r^2 + b^2 ) \\
&& + \beta r^2 (r^2 + b^2))
\Biggl(\frac{\partial}{\partial r}
\beta \Biggr)^2 \\
\kappa_2 & = & - \cos^2\theta ( a^2 - b^2)  + ( r^2 + b^2) .
\end{eqnarray*}

The  black  hole  temperature  $T$  can be  easily  computed  by  $T =
\kappa/{2   \pi}$.   The   pure   Kerr   5D   formula,  as  given   in
\cite{hawking2}, leads to:
\begin{eqnarray}
T = \frac{r_+^2 \Delta'}{4 \pi (r_+^2 +a^2) (r_+^2+ b^2)},
\end{eqnarray}
where $\Delta  = (r_+^2 + a^2)(r_+^2  + b^2)/r_+^2$ .

Fig.1 displays the pure Kerr  temperature  and  the  Kerr-Gauss-Bonnet
temperature.  As  expected, both values become very  close  for  large
masses. They differer by  about 5\% in the limit of very  small masses
for $\alpha=1$ in Planck units.

\begin{figure}
[htbp]
$$
\epsfxsize=8cm
\epsfbox{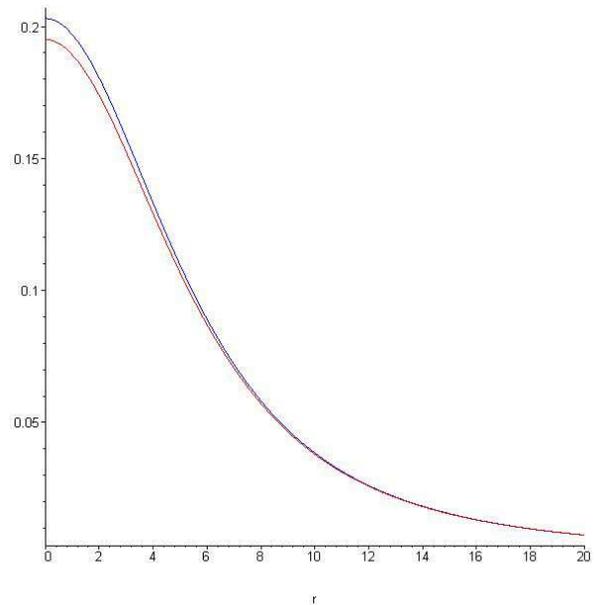}
$$
\caption{ Black hole  temperature  $T$ (y-axis, relative Plank values)
versus black hole size $r_+$ (x-axis, relative Plank  values) for pure
Kerr case (lower line) and Kerr-Gauss-Bonnet case (upper line).}
%\label{}
\end{figure}

\section{Discussion and conclusions}

If, as suggested  by geometrical arguments and by low-energy effective
superstring theories, Gauss-Bonnet gravity is a  realistic path toward
a full quantum theory  of  gravity, then Kerr-Gauss-Bonnet black holes
are  probably  among  the  most  important  objects to understand  the
physical basis of our World. This article has established the solution
of Einstein equations  in  the 5-dimensional Gauss-Bonnet theory. This
allows investigating into the details  the  physics  of  ``realistic''
spinning  black  holes,  both  from  a  pure theoretical  and  from  a
phenomenological  (in   the  framework  of  low  Planck-scale  models)
viewpoint.

Some improvements and developments  can  be foreseen. First, it should
be very welcome to obtain the same kind of solutions for any number of
dimensions. Unfortunately the method introduced in this article is not
easy to generalize and a specific  study should be made for each case.
Then, it would  be  interesting to  compute  the greybody factors  for
those black holes. Following the techniques  of  \cite{grain},  it  is
possible (although not straightforward) to obtain a numerical solution
as soon as the metric is  known, at least in the $\Lambda=0$ case. The
Kerr-Gauss-Bonnet-(Anti)-deSitter situation is more  intricate  as the
metric is nowhere flat, requiring  a  more  detailed investigation, as
suggested in \cite{kanti}.

\section{Addendum}

After the completion of this work, it was pointed out by N. Deruelle
that some analytical checks ({\it e.g.} on the trace of the field
equations) with Mathematica suggest that this solution is
not exact. It could mean that the Kerr-Gauss-Bonnet solution should be 
looked for in a wider class of solutions than Kerr-Schild ones.
Nevertheless, our results remain "heuristically" relevant as an approximation 
taking into account the accurate numerical checks.

\end{document}